\documentclass[twocolumn,showpacs,prl]{revtex4}

\usepackage{graphicx}%
\usepackage{dcolumn}
\usepackage{amsmath}
\usepackage{latexsym}

\begin {document}

\title
{
Small-world properties of the Indian Railway network
}
\author
{
Parongama Sen$^1$, Subinay Dasgupta$^1$, Arnab Chatterjee$^2$, P. A. Sreeram$^3$, 
G. Mukherjee$^{3,4}$ and S. S. Manna$^3$
}
\affiliation
{
$^1$Department of Physics, University of Calcutta, 
92 Acharya Prafulla Chandra Road, Kolkata-700009, India\\
$^2$TCMPD, Saha Institute of Nuclear Physics, Block-AF, 
Sector-I, Salt Lake, Kolkata-700064, India\\
$^3$Satyendra Nath Bose National Centre for Basic Sciences 
Block-JD, Sector-III, Salt Lake, Kolkata-700098, India\\
$^4$Bidhan Chandra College, Asansol 713304, Dt. Burdwan, 
West Bengal, India
}

\begin{abstract}
Structural properties of the Indian Railway network is studied in the light of
recent investigations of the scaling properties of different complex networks.
Stations are considered as `nodes' and an arbitrary pair of stations
is said to be connected by a `link' when at least one train stops at both stations.
Rigorous analysis of the existing data shows that the Indian Railway 
network displays small-world properties. We define and estimate several
other quantities associated with this network.

\end{abstract}
\maketitle

      Given a chance, how would we have possibly organized our train travel? People
   dislike to change trains to reach their destinations. Therefore an extreme
   possibility would be to run a single train passing through all stations in the 
   country so that no change of train is needed at all! An obvious disadvantage in this
   strategy is that the average distance between the stations become very large and
   so also the time needed for travel. The other limiting situation would be, to run
   a train between any pair of neighbouring stations and try to
   travel along the minimal paths. This requires a change of train at every station,
   which is also clearly not economically viable. Railway networks in no country in the world 
   follow either of the two ways, actually they go mid-way. Like any other transport
   system the main motivation of railways is to be fast and economic. To achieve it,
   railways run simultaneously many trains, covering short as well as long routes
   so that a traveller does not need to change more than only a few trains to reach
   any arbitrary destination in the country.

      In this paper we analyse the structure of the Indian Railway network (IRN). This 
   is done in the context of recent investigations of the scaling properties of several 
   complex networks e.g., social, biological, computational networks \cite {Small} etc. 
   Identifying the stations as nodes of the network and a train which stops at any two 
   stations as the link between the nodes we measure the average distance between an 
   arbitrary pair of stations and 
   find that it depends only logarithmically on the total number of stations in the country. 
   While from the network point of view this implies the small-world nature of the railway 
   network, in practice a traveller has to change only few trains to reach an arbitrary 
   destination. This implies that over years, the railway network has been evolved with the 
   sole aim in mind to make it fast and economic, eventually its structure has become a 
   small-world network \cite {WS}.

      The structure and properties of several social, biological and computational networks 
   like the World-wide web (WWW) \cite {web}, network of the Internet structure \cite 
   {Faloutsos}, neural networks \cite {neural}, collaboration network \cite {collab} etc. 
   are being studied recently with much interest. In general a network has a number of `nodes' 
   and some `links' connecting different pairs of nodes. Typically the following quantities 
   are defined to characterize a network of $N$ nodes: (i) the diameter is the maximum
   distance between an arbitrary pair of nodes (ii) the 
   clustering coefficient ${\cal C}(N)$ is the average fraction of connected triplets (iii) 
   the probability distribution $P(k)$ that an arbitrarily selected node has the degree $k$ 
   i.e. this node is linked to $k$ other nodes. 

\begin{figure}[top]
\begin{center}
\includegraphics[width=6.5cm]{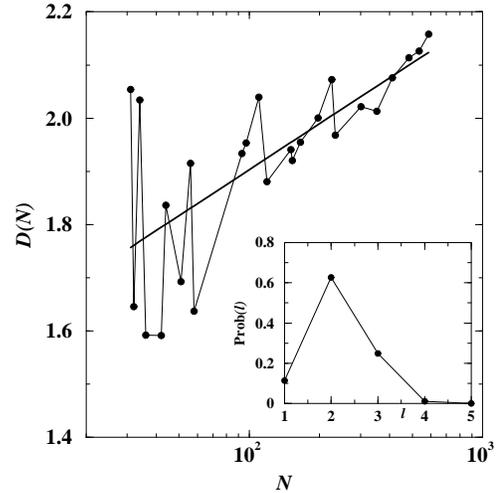}
\end{center}
\caption{
The variation of the mean distance ${\cal D}(N)$ of 25 different subsets of IRN having
different number of nodes $(N)$. The whole range is fitted with a function like
${\cal D}(N) = A+B\log(N)$ where $A \approx 1.33$ and $B \approx 0.13$.
The inset shows the distribution Prob$(\ell)$ of the shortest path lengths $\ell$
on IRN. The lengths varied to a maximum of only five link lengths
and the network has a mean distance ${\cal D}(N) \approx 2.16$. 
}
\end{figure}

      Watts and Strogatz \cite{WS} proposed a model of small-world 
   network (SWN) in the context of various social and biological networks. They argued that
   SWNs must have small diameters which grow as $\ln N$ like random
   networks but should have large values of the clustering coefficients ${\cal C}(N) \sim 1$
   like regular networks. On the other hand the scale-free networks (SFN) are characterized
   by the power law decay of the degree distribution function: $P(k) \sim k^{-\gamma}$. 
   It was observed later that the degree distributions of nodes for two very important 
   networks e.g., World-wide web \cite {web} which is a network of web-pages and the 
   hyperlinks among various pages and the Internet network \cite {Faloutsos} of routers 
   or autonomous systems have scale-free property. Barab\'asi and Albert (BA) proposed a 
   model for SFN which grows from an initial set of nodes and at every time step some 
   additional nodes are introduced which are randomly connected to the previous nodes with the 
   linear attachment probabilities \cite {barabasi}. All scale-free networks are 
   believed to display small-world properties while a small-world network is not necessarily 
   scale-free.

\begin{figure}[top]
\begin{center}
\includegraphics[width=6.5cm]{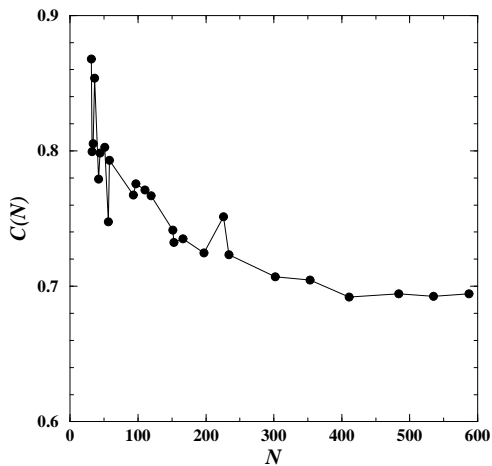}
\end{center}
\caption{
Variation of the clustering coefficient ${\cal C}(N)$ of 25 different subsets of IRN having
different number of nodes $N$. Starting from a somewhat higher value at small number
of nodes, the clustering coefficient decreases slowly on increasing $N$ and finally
saturates at 0.69.
}
\end{figure}

      Networks defined on the Euclidean space have also generated much interests in recent times.
   Internet, transport systems, postal networks etc. are naturally defined on two-dimensional space.
   In these generalised networks the attachment probabilities depend jointly on the nodal
   degrees as well as the lengths of the links \cite{euclid1,euclid2}. 

      A railway network is one of the most important examples of transport systems. The
   very complex topological structures of railway networks have attracted the 
   attention of researchers in many different contexts. For example the fractal nature 
   of the structure of railway networks was studied by Benguigui \cite {Benguigui}.
   Very recently the efficiency of Boston subway network has been studied where
   a new measure for such networks has been proposed \cite {boston}.

\begin{figure}[top]
\begin{center}
\includegraphics[width=6.5cm]{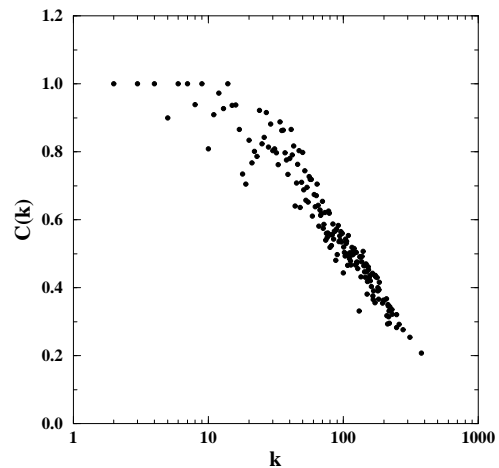}
\end{center}
\caption{
The variation of the clustering coefficient ${\cal C}(k)$
against the degree $k$ for the IRN
indicates a logarithmic decay for large $k$.
}
\end{figure}

      Our scheme is to associate first a representative graph $G_N$ with the IRN 
   of $N$ stations in the following way. Here the stations represent the `nodes' 
   of the graph, whereas two arbitrary stations are considered to be connected 
   by a `link' when there is at least one train which stops at both the stations. 
   These two stations are considered to be at unit distance of separation irrespective 
   of the geographical distance between them. Therefore the shortest distance 
   $\ell_{ij}$ between an arbitrary pair of
   stations $s_i$ and $s_j$ is the minimum number of different trains one needs to board 
   on to travel from $s_i$ to $s_j$. Thus $\ell_{ij}=1$ implies that there is at least one 
   train which stops at both $s_i$ and $s_j$. Similarly, $\ell_{ij}=2$ 
   implies that there is no train which stops at both $s_i$ and $s_j$ and one has to 
   change the train at least once in some intermediate station to board the
   second train to reach $s_j$. With this definition, if the trains $t_1$, $t_2$, 
   $\cdots$ $t_n$ pass
   through a station $s_i$, then all the stations through which these $n$ trains
   pass are unit distance away from $s_i$ and are considered as first neighbours of $s_i$.
   Consequently, the number $k_i$ of such stations is the degree of the node $s_i$.

      Indian Railway network is a densely populated network of more than 8000 stations 
   where the number of trains plying in this network is of the order of 10000 \cite {IRsite}.
   However, we collected the data of IRN on a coarse-grained level following the
   recent Indian Railways time table `Trains at a Glance' \cite {Timetable}
   containing the important trains and stations in India. This table contains a total 
   of $L=579$ trains covering $N=587$ stations in a total of 86 tables. A grand 
   rectangular matrix ${\cal G}(N,L)$ is then constructed such that the $ij$-th 
   element of this matrix is 1 if the train $j$ stops at the station $i$, otherwise 
   this element is zero. A second matrix ${\cal T}(0:N,N)$ is also
   constructed where the degree $k_i$ of the station $i$ is stored
   at the element ${\cal T}(0,i)$ and the serial numbers of the $k_i$ neighbours of
   $i$ are stored at the locations ${\cal T}(j,i), j=1,k_i$, rest of the elements
   being zero. We define and estimate the following quantities for the IRN.

\begin{figure}[top]
\begin{center}
\includegraphics[width=6.5cm]{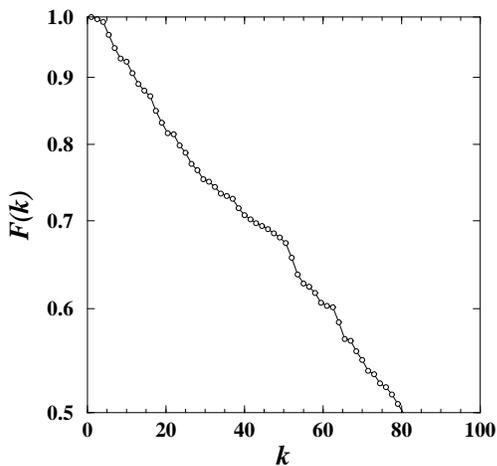}
\end{center}
\caption{
The cumulative degree distribution $F(k)$ of the IRN with the degree $k$ is
plotted on the semi-logarithmic scale. 
}
\end{figure}

      Since $G_N$ is a connected graph, there are $N(N-1)/2$ distinct shortest 
   paths among the $N$ stations. We calculate the probability distribution of 
   the shortest path lengths Prob${(\ell)}$. The shortest path lengths are calculated
   using a burning algorithm \cite {Herrmann} and using the matrix ${\cal T}$. 
   In this algorithm the fire starts from an arbitrary node $i$, and burns this 
   node at time $t=0$. At time $t=1$ the fire burns all $k_i$ neighbours of $i$. 
   At time $t=2$ all unburnt neighbours of $k_i$ nodes are burnt and so on. The 
   burning time of a node is the length of the shortest path of that node from 
   the node $i$. This calculation has been repeated for all $N$ nodes to get 
   $N(N-1)/2$ shortest distances. In Fig. 1 inset we plot this distribution which 
   goes to a maximum of $\ell=5$ implying that one needs to change at most four 
   trains to reach any station from any station in India on the coarse-grained 
   level. Similarly the distribution has a peak at $\ell=2$ implying that one 
   can go to the majority of stations in India by changing train only once. 
   In the graph theory the diameter of a graph is measured by the maximum
   distance between the pairs of nodes. Therefore according to this definition the diameter of our
   network is exactly equal to 5. However the average shortest path between an arbitrarily selected pair of nodes
   which we call as the mean distance ${\cal D}(N)$
   is also a measure of the topological size of the graph and have been
   used by many authors to measure the size of networks as described in \cite {barabasi}. We therefore
   measure the mean distance ${\cal D}(N)$ of the railway network of $N$ stations as the average shortest 
   distance $\langle {\ell}_{ij} \rangle$ between an arbitrary pair of stations 
   $s_i$ and $s_j$. We obtain ${\cal D}(N) \approx 2.16$ for this network.

      It is desirable to see how ${\cal D}(N)$ varies with $N$
   \cite {amaral}. Since we have the 
   data of a single railway network, we divide the whole IRN into 25 different 
   subsets consisting of trains and stations of 10 different states, 7 different 
   combinations of states, 7 different railway zones and the whole IRN. As a result 
   we obtained 25 data points (though they are not necessarily non-overlapping samples), 
   reflecting the nature of variation of ${\cal D}(N)$
   with $N$. In Fig. 1 we plot this data on a semi-log scale and though there is 
   some wild fluctuations for small values of $N$, for large values of $N$ the linear 
   behaviour is quite apparent. The whole range is fitted with ${\cal D}(N) = A+B\log(N)$
   where $A \approx 1.33$ and $B \approx 0.13$.

\begin{figure}[top]
\begin{center}
\includegraphics[width=6.5cm]{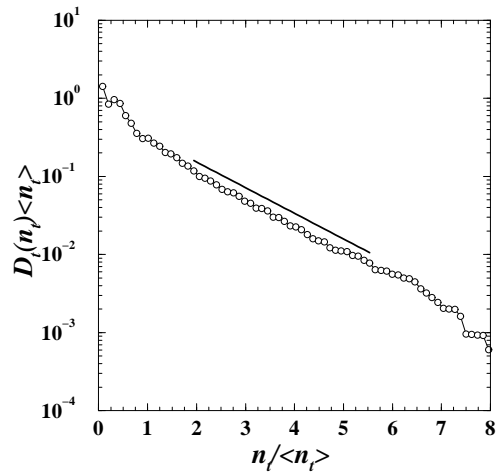}
\end{center}
\caption{
Scaled probability distribution $D_t(n_t)$ for an arbitrary station through which $n_t$ trains
pass where $\langle n_t \rangle \approx 12.06$.
Binned data is presented through the circles connected by lines 
which fits best to an exponential form: $D_t(n_t)\langle n_t \rangle = a\exp(-bx)$ 
with $x=n_t/\langle n_t \rangle$, $a \approx 0.47$ and $b \approx 0.75$.
}
\end{figure}

      The clustering coefficient ${\cal C}(N)$ is defined in the following way.
   Let the subgraph $G_i$ consisting of the neighbours of $s_i$ i.e., $(s_1, s_2, 
   s_3, \cdots, s_{k_i})$ have $E_i$ links among them. Then the clustering co-efficient 
   ${\cal C}_i$ of the node $i$ is $2E_i/k_i(k_i-1)$ and that of the whole network is 
   ${\cal C}_i = \langle {\cal C}_i \rangle$. 
   A direct measure of the clustering co-efficient of the
   whole IRN gives: ${\cal C} \approx 0.69$ (Fig. 2). 
The high value of the clustering coefficient is explained in the following
way. The number $n_s$ of stations in which a particular train stops are all
at unit distance from one another on the network and therefore form an
$n_s$-clique. Therefore if only one train stops at some station $i$ then
$C_i=1$. When two trains stop at the station $i$ and the sets $n_s(1)$ and $n_s(2)$ of stations
covered by these two trains are different, $C_i$ is in general smaller than 1.
However there may be other trains which do not stop at $i$ but stop at
the stations which are not in both $n_s(1)$ and $n_s(2)$. These trains
enhance the value of $C_i$.
The value of ${\cal C} \approx 0.69$ is compared with a corresponding 
   random graph network having the same number of vertices and edges as in IRN
   with the edges distributed randomly. It is found that the number of edges in IRN
   is 19603. If these edges are distributed randomly within the maximum possible 
   edges on a graph of $N$=587 nodes the the clustering coefficient should be 
   19603/$[N(N-1)/2] \approx 0.113$ which is the same as Prob(1). We also compute a 
   modified clustering coefficient ${\cal C}_o$ by counting in $E_i$ only those links 
   in the subgraph $G_i$ which pass through the node $i$. We obtained a value 
   ${\cal C}_o \approx 0.55$ for the IRN.

   Recently, the study of the clustering coefficient as a
function of the degree of the node of some real-world network
has shown an interesting feature \cite {Ravasz}
${\cal C}(k)$, defined as the clustering coeffcient of the
node with degree $k$, shows a decrease (apparently a power law decay) with
$k$ in several networks like the actor, language or world-wide-web networks.
However in the network of internet at the router level or power grid
network of the Western US, ${\cal C}(k)$ was found to be more
or less a constant. In the IRN also, we find that ${\cal C}(k)$ (Fig. 3)
remains at a constant value close to unity for small $k$ and shows a
logarithmic decay at larger values of $k$. In all these real-world networks
where ${\cal C}(k)$ remains more or less a constant, 
the nodes are linked by physical connections
which may be responsible for this common feature. However, in this 
context it should also
be mentioned that the scale-free Barab\'asi-Albert network \cite{barabasi}
also predicts ${\cal C}(k) \propto k^0$ and  ${\cal C}(N) \propto N^{-0.75}$.
In the IRN, although ${\cal C}(N)$ shows a decrease with $N$, it is apparently
much slower than a power law.

      The degree distribution of the network, that is, the distribution of the 
   number of stations $k$ which are connected by direct trains to an arbitrary station
   is denoted by $P(k)$. We plot the cumulative degree distribution 
   $F(k) = \int_k^{\infty} P(k)dk $ using a semi-log scale in Fig. 4 for the whole IRN. 
   We see that $F(k)$ approximately fits to an exponentially decaying distribution
   $F(k) \sim exp(-\alpha k)$ with $\alpha$= 0.0085.
   
\begin{figure}[top]
\begin{center}
\includegraphics[width=6.5cm]{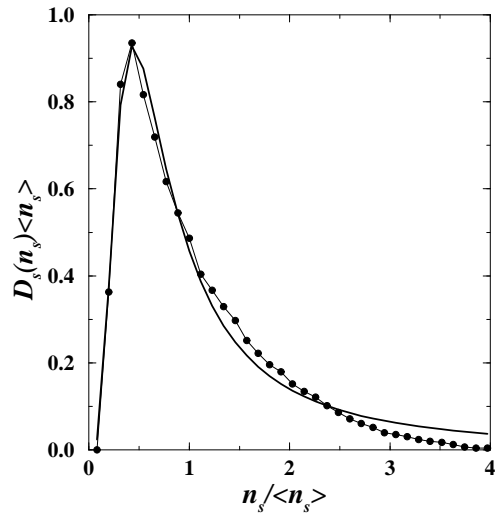}
\end{center}
\caption{
Scaled probability distribution $D_s(n_s)$ for an arbitrary train passing through $n_s$ stations
where $\langle n_s \rangle \approx 12.37$.
Binned data is presented through the black dots connected by lines 
which fits best to the form: $D_s(n_s)\langle n_s \rangle = ax^4/(x^2+b)^3$ 
with $x=n_s/\langle n_s \rangle$, $a \approx 0.6$ and $b \approx 0.096$.
}
\end{figure}

      We also calculated the distribution $D(n_t)$ of the number of trains $n_t$ which 
   stop at an arbitrary station. This is plotted in Fig. 5 on a semi-log scale after
   scaling by the average number of trains $\langle n_t \rangle \approx 12.06$ along both the abscissa 
   and the ordinate. The data is binned as before and is fitted to an exponential form:
   $D_t(n_t)\langle n_t \rangle = a\exp(-bx)$ with $x=n_t/\langle n_t \rangle$, 
   $a \approx 0.47$ and $b \approx 0.75$.

      The distribution $D(n_s)$ of the number of stations through which an arbitrary train 
   passes is plotted in Fig. 6. The data is scaled by the average number of stations
   $\langle n_s \rangle \approx 12.37 $ along both the abscissa and the ordinate.
   The $D(n_s)$ grows very fast at the beginning, reaches a maximum and then decays
   to zero. A numerical fit to a functional form like $D_s(n_s)\langle n_s \rangle = 
   ax^4/(x^2+b)^3$ with $x=n_s/\langle n_s \rangle$, $a \approx 0.6$ and $b \approx 0.096$
   turns out to be reasonably good.

      We also measure the connectivity correlation of IRN following the
   works of \cite {Pastor}. Let $F(k'|k)$ denote the conditional probability that
   a node of degree $k$ has a neighbour of degree $k'$. Then to see how the 
   nodes of different degrees are correlated we measure the average degree 
   $\langle k_{nn}(k) \rangle = \Sigma_{k'}k'F(k'|k) $ 
   of the subset of nodes which are all neighbours to a particular node of degree $k$.
   In general this average has a variation like $\langle k_{nn}(k) \rangle \sim k^{-\nu}$
   where a non-zero $\nu$ reflects a non-trivial correlation among the nodes
   of the network. We calculated $\langle k_{nn}(k) \rangle$ for IRN and plotted it
   in Fig. 7 on a double logarithmic scale. Almost over a decade the  $\langle k_{nn}(k) \rangle$
   remains same on the average and is independent of $k$, indicating the
   absence of correlations among the nodes of different degrees.

A more sensitive measure for the degree correlations was proposed in \cite {Newman}.
Newman has defined a degree-degree correlation function $r$ which
measures whether a vertex of high degree at one end of a link prefers
a vertex of {\em high} degree (``assortative mixing'', $r>0$) or {\em low} 
degree (``disassortative mixing'' $r<0$) at the other end. It has been
observed that social networks are assortative and technological and biological 
networks are disassortative. We have measured for IRN the normalized 
correlation function following \cite {Newman} and found its values to be 
$r$ = -0.033. This indicates that the IRN is of {\em disassortative} nature, i.e. rich
vertices at one end of a link show some preference towards poor vertices
at the other end, and vice versa.

\begin{figure}[top]
\begin{center}
\includegraphics[width=6.5cm]{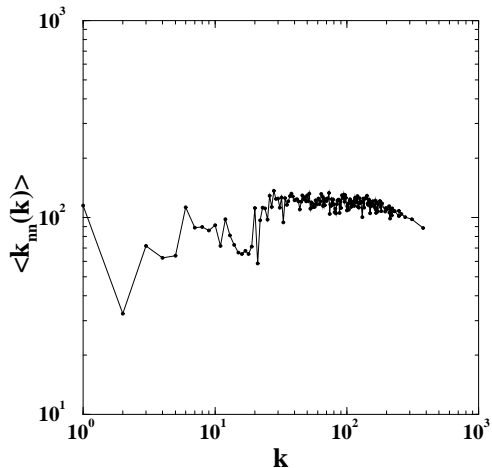}
\end{center}
\caption{
The variation of the average degree $\langle k_{nn}(k) \rangle$ of the neighbours of a node of
degree $k$ with $k$. After some initial fluctuations,  $\langle k_{nn}(k) \rangle$ remains
almost same over a decade around $k$ = 30 to 300 indicating absence of correlations 
among the nodes of different degrees.
}
\end{figure}

      To summarize, we investigated the structural properties of the Indian Railway
   network to see if some of the general scaling behaviour obtained for many complex 
   networks in recent times may also be present in IRN. While nodes of the network
   are evidently the stations, the links are defined as the pairs of stations 
   communicated by single trains. With such a definition of link, the mean distance of the
   network is a measure of how good is the connectivity of the network.
   Indeed, we observed that the mean distance of IRN varies logarithmically with the
   number of nodes with a high value of the clustering coefficient. This implies that
   IRN behaves like a small-world network, which we believe should be typical of 
   the railway network of any other country, which we are unable to study at present for 
   unavailability of data.

   We like to thank I. Bose for constantly encouraging us to work on this
   problem and also to S. Goswami for suggesting ref. [13]. PS 
   acknowledges financial support from DST grant SP/S2-M11/99.
   GM acknowledges hospitality in the S. N. Bose National Centre.

\end{document}